\documentclass[12pt]{article}
\usepackage{amsfonts}
\usepackage{amsmath}

\input{tcilatex}
\begin{document}

\title{Chaotifiying continuous-time systems by symmetry}
\author{Zeraoulia Elhadj$^{1}$, J. C. Sprott$^{2}$ \\
%EndAName
$^{1}$Department of Mathematics, University of T\'{e}bessa, (12002), Algeria%
\\
E-mail: zeraoulia@mail.univ-tebessa.dz and zelhadj12@yahoo.fr\\
$^{2}$ Department of Physics, University of Wisconsin, Madison, WI 53706,
USA \\
E-mail: sprott@physics.wisc.edu}
\maketitle

\begin{abstract}
In this letter we present a method of constructing dynamical systems with
any preassigned number of equilibria by adding symmetry to another system
with at least one equilibrium point. If the resulting system is chaotic, we
call this procedure \textit{chaotification by symmetry} since the resulting
system is chaotic with many symmetrical equilibrium points.
\end{abstract}



\section{Introduction}

There are several well-known situations involving computation of equilibria,
such as Nash equilibria and market equilibria. For these two cases, the
equilibrium is the result of individual agents acting on their own in a
noncompetitive but selfish manner. See [10] for more details. On the other
hand, the existence of many equilibria in a system makes its dynamics more
complex and shows some special structure. Examples include the well-known
multi-scroll attractors [1-2-4-5-6-7 and references therein], chaotic
attractors with multiple-merged basins of attraction [3], scroll grid
attractors [8], and $2n$-wing and $n\times m$-wing Lorenz-like attractors
[12-13].

In [14] a methodology was presented for constructing some simple chaotic
systems with any preassigned number of equilibria by adding symmetry to a
simple 3-D quadratic system with only one stable equilibrium.

In this letter, we generalize the above method to any $m$-dimensional system
by introducing a singular transformation to generate a system with an
explicit formula and $n$ equilibrium points. The importance of this result
is that systems with many equilibria are generally more complex and display
chaotic attractors with special structure. See for example [9].

\section{Generating $m$-dimensional systems with $n$ equilibrium points}

In this section, we present our method to generate an $m$-dimensional system
with exactly $n$ equilibrium points. Consider the following $m$-dimensional
system:%
\begin{equation}
\left\{ 
\begin{array}{c}
x_{1}^{\prime }=f_{1}\left( x_{1},x_{2},...,x_{m}\right) \\ 
x_{2}^{\prime }=f_{2}\left( x_{1},x_{2},...,x_{m}\right) \\ 
.... \\ 
x_{m}^{\prime }=f_{m}\left( x_{1},x_{2},...,x_{m}\right)%
\end{array}%
\right.
\end{equation}%
and consider the following coordinate transformation%
\begin{equation}
\left\{ 
\begin{array}{c}
u_{1}=\left( \sqrt{x_{1}^{2}+x_{m}^{2}}\right) ^{\frac{1}{n}}\cos \left( 
\frac{1}{n}\arccos \left( \frac{x_{1}}{\sqrt{x_{1}^{2}+x_{m}^{2}}}\right)
\right) \\ 
u_{j}=x_{j},\text{ for }j=2,3,...,m-1 \\ 
u_{m}=\left( \sqrt{x_{1}^{2}+x_{m}^{2}}\right) ^{\frac{1}{n}}\sin \left( 
\frac{1}{n}\arccos \left( \frac{x_{1}}{\sqrt{x_{1}^{2}+x_{m}^{2}}}\right)
\right)%
\end{array}%
\right.
\end{equation}%
where $n$ is the number of desired equilibria. Transformation (2) can add a $%
\left\{ x_{j,}j=2,3,...,m-1\right\} $-axis rotation symmetry, $%
%TCIMACRO{\U{211d} }%
%BeginExpansion
\mathbb{R}
%EndExpansion
_{\left\{ x_{j,}j=2,3,...,m-1\right\} }(\frac{2}{n}\pi )$, to the original
system.

The first and the last relations of the transformation (2) are deduced from
the relation $x_{1}+ix_{m}=\left( u_{1}+iu_{m}\right) ^{n}$ in the complex
plane, where $u_{1}+iu_{m}$ is the unknown variable$.$ Let $%
x_{1}+ix_{m}=r\left( \cos \theta +i\sin \theta \right) ,$ where $r=\sqrt{%
x_{1}^{2}+x_{m}^{2}},$ $\cos \theta =\frac{x_{1}}{\sqrt{x_{1}^{2}+x_{m}^{2}}}%
,$ and $\sin \theta =\frac{x_{n}}{\sqrt{x_{1}^{2}+x_{m}^{2}}}.$ Let $%
u_{1}+iu_{m}=s\left( \cos \alpha +i\sin \alpha \right) ,$ where $s=\sqrt{%
u_{1}^{2}+u_{m}^{2}},$ $\cos \alpha =\frac{u_{1}}{\sqrt{u_{1}^{2}+u_{m}^{2}}}%
,$ and $\sin \alpha =\frac{u_{n}}{\sqrt{u_{1}^{2}+u_{m}^{2}}}.$ By
comparaison, we have $r=s^{n}$ and $\alpha =\frac{\theta +2k\pi }{n}%
,k=0,1,...,n-1.$ Here we can choose $\theta =\arccos \left( \frac{x_{1}}{%
\sqrt{x_{1}^{2}+x_{m}^{2}}}\right) $ ($k=0$). The other cases are similar,
but are complex in their analysis. Finally, we can obtain the formulas for $%
u_{1}$ and $u_{m}$ as in (2). On the other hand, we have $x_{1}=s^{n}\cos
n\alpha $ and $x_{m}=s^{n}\sin n\alpha .$ Hence we have%
\begin{equation}
\left\{ 
\begin{array}{c}
x_{1}=\left( \sqrt{u_{1}^{2}+u_{m}^{2}}\right) ^{n}\cos \left( n\arccos
\left( \frac{u_{1}}{\sqrt{u_{1}^{2}+u_{m}^{2}}}\right) \right) =\varphi
_{1}\left( u_{1},u_{m}\right) \\ 
x_{j}=u_{j},\text{ for }j=2,3,...,m-1 \\ 
x_{m}=\left( \sqrt{u_{1}^{2}+u_{m}^{2}}\right) ^{n}\sin \left( n\arccos
\left( \frac{u_{1}}{\sqrt{u_{1}^{2}+u_{m}^{2}}}\right) \right) =\varphi
_{2}\left( u_{1},u_{m}\right)%
\end{array}%
\right.
\end{equation}%
By differentiating this equation with respect to $t$, we have%
\begin{equation}
\left\{ 
\begin{array}{c}
u_{1}^{\prime }=\frac{\left( \frac{\partial \varphi _{1}}{\partial u_{1}}%
\frac{\partial \varphi _{2}}{\partial u_{m}}-\frac{\partial \varphi _{2}}{%
\partial u_{1}}\frac{\partial \varphi _{1}}{\partial u_{m}}\right) f_{1}-%
\frac{\partial \varphi _{1}}{\partial u_{m}}\left( \frac{\partial \varphi
_{1}}{\partial u_{1}}f_{m}-\frac{\partial \varphi _{2}}{\partial u_{1}}%
f_{1}\right) }{\frac{\partial \varphi _{1}}{\partial u_{1}}\left( \frac{%
\partial \varphi _{1}}{\partial u_{1}}\frac{\partial \varphi _{2}}{\partial
u_{m}}-\frac{\partial \varphi _{2}}{\partial u_{1}}\frac{\partial \varphi
_{1}}{\partial u_{m}}\right) } \\ 
u_{j}^{\prime }=x_{j}^{\prime }=f_{j}\left( \varphi _{1}\left(
u_{1},u_{m}\right) ,u_{2},u_{3},...,\varphi _{2}\left( u_{1},u_{m}\right)
\right) ,\text{ for }j=2,3,...,m-1 \\ 
u_{m}^{\prime }=\frac{\frac{\partial \varphi _{1}}{\partial u_{1}}f_{m}-%
\frac{\partial \varphi _{2}}{\partial u_{1}}f_{1}}{\frac{\partial \varphi
_{1}}{\partial u_{1}}\frac{\partial \varphi _{2}}{\partial u_{m}}-\frac{%
\partial \varphi _{2}}{\partial u_{1}}\frac{\partial \varphi _{1}}{\partial
u_{m}}}%
\end{array}%
\right.
\end{equation}%
where $f_{k}=f_{k}\left( \varphi _{1}\left( u_{1},u_{m}\right)
,u_{2},u_{3},...,\varphi _{2}\left( u_{1},u_{m}\right) \right) $ for $k\in
\left\{ 1,m\right\} $. Here we use the facts that $\left( \varphi _{1}\left(
u_{1},u_{m}\right) \right) ^{\prime }=\frac{\partial \varphi _{1}}{\partial
u_{1}}u_{1}^{\prime }+\frac{\partial \varphi _{1}}{\partial u_{m}}%
u_{m}^{\prime }$ and Transformation (2) is singular and not defined at the
points $\left( 0,x_{2},x_{3},...,0\right) $. Hence the two systems (1) and
(4) are not globally but only locally topologically equivalent. Equation (4)
is well defined if and only if $\frac{\partial \varphi _{1}}{\partial u_{1}}%
\neq 0$ and $\frac{\partial \varphi _{1}}{\partial u_{1}}\frac{\partial
\varphi _{2}}{\partial u_{m}}-\frac{\partial \varphi _{2}}{\partial u_{1}}%
\frac{\partial \varphi _{1}}{\partial u_{m}}\neq 0.$ But these relations are
not true for all variables $u_{1}$ and $u_{m}$ since, for example, $\frac{%
\partial \varphi _{1}\left( u_{1},u_{m}\right) }{\partial u_{1}}$ can vanish
for the set of points satisfying $\tan \left( n\arccos \frac{u_{1}}{\sqrt{%
u_{1}^{2}+u_{m}^{2}}}\right) =\frac{-\allowbreak u_{1}}{\left\vert
u_{m}\right\vert }.$

The equilibrium points of the new system (4) are the real solutions of the
algebraic equations $u_{j}^{\prime }=0$ for all $j=1,2,...,m.$ Thus we have%
\begin{equation}
f_{j}\left( \varphi _{1}\left( u_{1},u_{m}\right) ,u_{2},u_{3},...,\varphi
_{2}\left( u_{1},u_{m}\right) \right) =0
\end{equation}%
for all $j=1,2,...,m$. Assume that the original system (1) has at least one
equilibrium point $\left( a_{1},a_{2},...,a_{m}\right) $. Then let $\left(
b_{1},b_{2},...,b_{m}\right) $ be an equilibrium point of the new system
(4). From (3) we have%
\begin{equation}
\left\{ 
\begin{array}{c}
a_{1}=\left( \sqrt{b_{1}^{2}+b_{m}^{2}}\right) ^{n}\cos \left( n\arccos
\left( \frac{b_{1}}{\sqrt{b_{1}^{2}+b_{m}^{2}}}\right) \right) \\ 
b_{j}=a_{j},\text{ for }j=2,3,...,m-1 \\ 
a_{m}=\left( \sqrt{b_{1}^{2}+b_{m}^{2}}\right) ^{n}\sin \left( n\arccos
\left( \frac{b_{1}}{\sqrt{b_{1}^{2}+b_{m}^{2}}}\right) \right)%
\end{array}%
\right.
\end{equation}%
Firstly, from the first and the last equation of (6), we have $\sqrt{%
b_{1}^{2}+b_{m}^{2}}=\left( \sqrt{a_{1}^{2}+a_{m}^{2}}\right) ^{\frac{1}{n}%
}. $ Secondly, in order to obtain $n$ values for $b_{1},$ we can assume that 
$a_{1}=0,$ (but $a_{m}\neq 0,$ otherwise the transformation is not defined)
that is $\cos \left( n\arccos \left( \frac{b_{1}}{\left( \left\vert
a_{m}\right\vert \right) ^{\frac{1}{n}}}\right) \right) =T_{n}\left( \frac{%
b_{1}}{\left( \left\vert a_{m}\right\vert \right) ^{\frac{1}{n}}}\right) =0,$
where $T_{n}$ is the Chebyshev polynomial of the first kind (well defined
here since $\frac{b_{1}}{\sqrt{b_{1}^{2}+b_{m}^{2}}}\in \left( -1,1\right) $%
) which has $n$ simple roots (multiplicity here is zero) of the form $\rho
_{k}=\frac{b_{1}^{\left( k\right) }}{\left( \left\vert a_{m}\right\vert
\right) ^{\frac{1}{n}}}=\cos \left( \frac{\left( 2k-1\right) \pi }{2n}%
\right) ,$ for all $k=1,...,n,$ in the interval $\left( -1,1\right) $. Hence 
$b_{1}$ has $n$ different values of the form $b_{1}^{\left( k\right)
}=\left( \left\vert a_{m}\right\vert \right) ^{\frac{1}{n}}\cos \left( \frac{%
\left( 2k-1\right) \pi }{2n}\right) ,$ for all $k=1,...,n.$ The last
equation of (6) and the fact that $\theta =\arccos \left( \frac{x_{1}}{\sqrt{%
x_{1}^{2}+x_{m}^{2}}}\right) =$ $\arcsin \left( \frac{x_{n}}{\sqrt{%
x_{1}^{2}+x_{m}^{2}}}\right) $ implies that $a_{m}=\left\vert
a_{m}\right\vert \sin \left( n\arcsin \left( \frac{b_{m}}{\left( \left\vert
a_{m}\right\vert \right) ^{\frac{1}{n}}}\right) \right) .$ Without loss of
generality, we can assume that $a_{m}>0$. Hence $b_{m}=a_{m}^{\frac{1}{n}%
}\sin \frac{\pi }{2n}.$ Thus the $n$ equilibrium points of the new system
(4) are of the form%
\begin{equation}
\left\{ 
\begin{array}{c}
b_{1}^{\left( k\right) }=a_{m}^{\frac{1}{n}}\cos \left( \frac{\left(
2k-1\right) \pi }{2n}\right) ,k=1,...,n \\ 
b_{j}=a_{j},\text{ for }j=2,3,...,m-1 \\ 
b_{m}=a_{m}^{\frac{1}{n}}\sin \frac{\pi }{2n}%
\end{array}%
\right.
\end{equation}%
The above procedure can generate $m$-dimensional systems with $n$ known
equilibrium points. From equations (7), we remark that if the original
system (1) has $q$ equilibrium points, then the new system (4) has $nq$
equilibrium points. If the resulting system is chaotic, then we call this
procedure \textit{chaotification by symmetry} since the resulting system is
chaotic with many symmetrical equilibrium points as shown in [14] for some
elementary examples.

\section{Conclusion}

In this letter a method was presented for constructing dynamical systems
with any preassigned number of equilibria by adding symmetry to another $m$%
-dimensional system with at least one equilibrium point. The importance of
this result is that systems with many equilibria are generally more complex
and display chaotic attractors with special structures as described in the
current literature.

\end{document}